\documentclass[12pt]{article}
\usepackage{graphicx}

\begin{document}
\begin{titlepage}
%
%







\begin{center}
\centerline{The phase equilibrium in a Lennard Jones fluid:possible applications in astrophysics}
\end{center}
\begin{center}
\centerline{V.Celebonovic}			
\end{center}

\begin{center}
\centerline{Institute of Physics,University of Belgrade, Pregrevica 118,11080 Zemun-Belgrade,Serbia}
\centerline{vladan@ipb.ac.rs}
\end{center}




\vskip3mm






\begin{abstract}
Using standard thermodynamics and previous results of the author, this paper aims to discuss the conditions for phase equilibrium in a Lennard-Jones fluid. Possibilities of  astrophysical applications of the results obtained here are discussed to some extent.\footnote{to appear in Serbian Astron.Journal}.
\end{abstract}
\end{titlepage}


{


\section{INTRODUCTION}

In this paper we shall explore the conditions for phase equilibrium in a fluid consisting of neutral atoms or molecules. Apart from being an interesting task in pure statistical physics, work on this problem has several astrophysical motivations. Fluids consisting of neutral atoms and/or molecules exist in the interiors and atmospheres of the giant planets and their icy satellites, but also in the diffuse molecular clouds which are present in large numbers in galaxies. A new field of possible applications of work discussed in this paper are the neutron stars. Recent work shows that the inner crust of neutron stars consists of a lattice of spherical nuclei imersed in a sea of free neutrons with a background of electrons [1].  

According to http://www.exoplanet.eu, at the end of December 2011., $716$ planets around other stars were known. Of all these planets, $83$ objects have orbital semi-major axes between $0.9$ and $1.4$ astronomical units. On general grounds, it can be expected that this range of distances from a star corresponds to temperatures under which fluids consisting of neutral atoms and molecules can exist, which illustrates the planetological importance of the study of such a fluid. Water,as an example of a planetologically and biologically important molecule can exist as a liquid in a region of a planetary system in which the temperature is between $273 K$ and $373 K$. The upper and lower radii of this zone (the so called habitable zone) depend on the absolute luminosity of a star. It can be shown that for a main sequence star with solar absolute luminosity, the inner and outer radii of the habitable zone are $0.95$ AU and $1.37$ AU. An example of this calculation is avaliable at the adress http://www.planetarybiology.com.   

Examples of interesting early work on the atmospheres of the giant planets are [2],[3] or [4]. Results of 
studies of such a fluid are also important in studies of cold interstellar clouds and protoplanetary disks which,because of their low temperature, contain neutral atoms and/or molecules. For an example of a recent observational study of protoplanetary disks see [5].   

The first necessary step in attempting to model any kind of a physical system is to determine in some way the form of the interparticle potential in it. Logically, in order to have the system in equilbrium, this potential must be a combination of an attractive and a repulsive part. In the calculations to be discussed in this paper, the so called Lennard-Jones (LJ) potential will be used. This potential has the following analytical form
\begin{equation}
	u(r)=4\epsilon\left[(\frac{\sigma}{r})^{12}-(\frac{\sigma}{r})^{6}\right] 
\end{equation}
The symbol $\epsilon$ denotes the depth of the potential,while $\sigma$ is the diameter of the molecular "hard core". Obviously,$lim_{r\rightarrow0}u(r)=\infty$. It can simply be shown that $lim_{r\rightarrow\sigma}u(r)= 0$ and that $(\partial u(r)/\partial r)=0$ for $r_{min}=2^{1/6}\sigma$. The depth of the potential well is $u(r_{min})= - \epsilon$. The term with $r^{-12}$ corresponds to the repulsive part of this potential. It is now known,for example [6], that much better results could be obtained by approximating this term with a function of the form $A\exp({- Br})$ (the so called Born-Meyer term). However, the $12-6$ potential has the virtue of simplicity. 

The following section contains a brief resume of the basic notions and main previous results, while the subsequent two parts contain the calculations and a discussion of their possible applications in astrophysics. 

\section{THE BASIC NOTIONS}

Many systems, both natural and in laboratory, show a number of different phases, each of which can behave in a different way.  The obvious question is what are the conditions under which these different phases can exist in equilibrium. The number of phases which can coexist in equilibrium within a system can be determined by the Gibbs phase rule,well known in theromodynamics. Coexisting phases are in thermal and mechanical equilibrium and can exchange matter [7]. In practical terms,this means that the temperature, pressure and the chemical potentials of the phases must be equal. 
As shown recently in [8] ,the chemical potential of the Lennard Jones fluid is given by
\begin{equation}
	\mu=\mu_{ID}-b_{0}p_{ID}\sum_{n=0}^{\infty}\frac{1}{ n!}\Gamma(\frac{2n-1}{4})(\frac{\epsilon}{T})^{\frac{2n+1}{4}} 
\end{equation} 
where $\mu_{ID}$ is the chemical potential of the ideal gas, $p_{ID}$ is the pressure of the ideal gas,$b_{0}=2\pi\sigma^{3}N_{A}/3$, $N_{A}$ is Avogadro's number, and $\Gamma$ denotes the gamma function. 
The pressure and temperature of a $LJ$ fluid are related by the virial development of the equation of state (EOS):
\begin{equation}
	\frac{p v}{k_{B}T}= \sum_{l=1}^{\infty}a_{l}(T)(n \lambda^{3})^{l-1}
\end{equation}
All the symbols on the left side of this equation have their standard meanings,while on the right side $a_{l}$ are the so called virial coefficients, $\lambda$ is the thermal wavelength and $v$ is the inverse number density $v=V/N=1/n$ . The thermal wavelength is given by
\begin{equation}
\lambda=(\frac{2\pi\hbar^{2}}{m k_{B} T})^{1/2}	
\end{equation}

where $\hbar$ is Planck's constant divided by $2 \pi$ , $k_{B}$ Boltzmann's constant and $m$ the particle mass. Due to increasing complexity with increasing order, the virial developement is most often truncated at second order terms. The first virial coefficient is $1$, and the second coefficient for the $LJ$ potential is given by (for example [7]).See also [9]. 
\begin{equation}
	a_{2}(T^*)=b_{0}\sum_{j=0}^{\infty}\gamma_{j}(1/T^*)^{(2j+1)/4}  
\end{equation}
where
\begin{equation}
	\gamma_{j}=\frac{- 2^{j+1/2}}{4 j!} \Gamma(\frac{2j-1}{4}) 
\end{equation} 
and
\begin{equation}
	T^* = \frac{k_{B} T}{\epsilon}
\end{equation}
The values of the first few coefficients $\gamma_{j}$ are: $\gamma_{0}= 1.733$,$\gamma_{1}=- 2.564$, $\gamma_{2}= - 0.8665$, and the explicit expression for the second virial coefficient is:
\begin{eqnarray}
	a_{2}= \frac{2\pi N_{A}\sigma^{3}}{3}[1.733(\frac{\epsilon}{k_{B}T})^{1/4}-\nonumber\\
2.56369(\frac{\epsilon}{k_{B}T})^{3/4}-0.8665 (\frac{\epsilon}{k_{B}T})^{5/4}-\ldots]
\end{eqnarray}
As $b_{0}=2\pi\sigma^{3}N_{A}/3$ is always positive, it can easily be shown that $a_{2}\leq0$ for $(\frac{\epsilon}{k_{B}T})> 0.32175$. 

\section{THE CALCULATIONS}

The second order approximation to the virial developement is 
\begin{equation}
	p=n k_{B} T (1+a_{2}n\lambda^{3})
\end{equation}
where the coefficient $a_{2}$ is given by eq.(7). Denoting two phases of a system by indexes $1$ and $2$,and applying to them the condition of the equality of temperatures and the pressure, one gets the following expression for the basic condition for phase equilibrium: 
\begin{equation}
	n_{1}k_{B}T+n_{1}^{2}k_{B}a_{21}T\lambda_{1}^{3}= n_{2}k_{B}T+n_{2}^{2}k_{B}a_{22}T\lambda_{2}^{3}
\end{equation}
which can be transformed into the form
\begin{equation}
	T(n_{1}-n_{2}) + T(n_{1}^{2}a_{21}\lambda_{1}^{3}-n_{2}^{2}a_{22}\lambda_{2}^{3})=0
\end{equation}
Introducing $n_{1}-n_{2}=x$, this expression can be solved to give
\begin{equation}
	x_{1,2}= n_{1}+\frac{1}{2a_{22}\lambda_{2}^{3}} [1\pm[1+4a_{22}n_{1}\lambda_{2}^{3}(1+a_{21}n_{1}\lambda_{1}^{3})]^{1/2}]
\end{equation}

\subsection{Analysis} 

Equation (12) can be expressed in the following form:
\begin{equation}
	x_{1,2} = n_{1}+ A[1\pm(1+y)^{1/2}]
\end{equation}
where 
\begin{equation}
	A= \frac{1}{2a_{22}\lambda_{2}^{3}}
\end{equation}
and 
\begin{equation}
	y = 4a_{22}n_{1}\lambda_{2}^{3}(1+a_{21}n_{1}\lambda_{1}^{3})
\end{equation}
Because of physical considerations,$x$ must be real, which implies that $y \geq -1$. The special case of $x=0$ leads to an interesting result. By definition of $x$ this means that $n_{1}=n_{2}$, and by eq.(11) it implies that $\lambda_{1}/\lambda_{2}=(a_{22}/a_{21})^{1/3}$. Finally,eq.(4) leads to the result that $m_{2}/m_{1}=(a_{22}/a_{21})^{2/3}$. 

It can easily be shown that $y$ will be negative if the product $a_{21}\times a_{22}$ is negative and $a_{22}$ negative,which means that $a_{21}$ has to be positive. Using the analytical expression for $a_{2}$ we have shown that $a_{2}\leq0$ for $(\frac{\epsilon}{k_{B}T})> 0.32175$. One of the conditions for phase equilibrium is  $T_{1}=T_{2}=T$ which implies that $\epsilon_{1}< 0.32175 k_{B} T$ and $\epsilon_{2}> 0.32175 k_{B}T$. 

In view of possible astrophysical applications, note that these two inequalities combine a material parameter ($\epsilon$) and an astrophysical parameter ($T$). For example, for $H_{2}$ ,which is present in the atmospheres of the giant planets,$\epsilon=33.3 k_{B}$  (for example [10]).This means that $a_{22}<0$ for $T<104 K$, which is a realistic limit in the atmosphere of Jupiter (for example).      

Assuming that $y= - 1$, it follows from eq.(13) that
\begin{equation}
	x_{1,2} = n_{1} + A 
\end{equation}
which further means that 
\begin{equation}
n_{2}= - A. 
\end{equation}

The thermal wavelength is positive by definition, so eq.(17) is physically posible if $a_{22}< 0$. The sign of $a_{2}$ depends on the material as well as the pressure and temperature.  

If $y=0$ a simple calculation shows that 
\begin{equation}
	x_{1,2} = n_{1} + A[1\pm1]
\end{equation}
which leads to two possibilities; either
\begin{equation}
	n_{2} = - \frac{1}{a_{22} \lambda_{2}^{3}}
\end{equation} 
or 
\begin{equation}
	n_{2}=0
\end{equation}
The result expressed by eq.(20) describes a physically interesting situation. It shows that $y=0$ corresponds to the case in which the number density of one of the phases is zero. An astronomical example of such a situation could be the top layer of the atmosphere of a giant planet. 

The limiting case $y$$\rightarrow$ $+$ $\infty$ has potentially very interesting physical implications. If the temperature of the material is sufficiently high, $\lambda\rightarrow0$, $y\rightarrow0$ but $A\rightarrow\infty$. This means that $x_{1,2}\rightarrow\infty$,which implies that $n_{1}\rightarrow\infty$ and $n_{2}$ is infinite of lower order than $n_{1}$. Such a situation looks similar to what one could expect in the outer layer of a neutron star. 

A case which is also physically interesting is  $y$$\rightarrow$$+$ $\infty$ but $T\rightarrow0$. This corresponds to $\lambda\rightarrow\infty$ ,but $A\rightarrow0$. Finally, it follows that in this case $x_{1,2}=n_{1}$,which further implies that $n_{2}\rightarrow0$ - the same result as the one obtained for $y=0$.

The function of the form $[1+y]^{1/2}$,assuming $\left|y\right|<1$,can be developed in Taylor's series as
\begin{eqnarray}
[1+y]^{1/2} = \sum_{n=0}^{\infty}\frac{(-1)^{n}(2n)!}{(1-2n)(n!)^{2}4^{n}} y^{n}
\end{eqnarray}
Inserting eq.(21) in eq.(13) gives the following explicit form of the result for $x_{1,2}$
\begin{equation}
	x_{1,2}=n_{1}+A[1\pm\sum_{n=0}^{\infty}\frac{(-1)^{n}(2n)!}{(1-2n)(n!)^{2}4^{n}} y^{n}]
\end{equation}
where $y$ and $A$ are given by eqs.(14) and (15). 
Assuming $|y|<1$ and taking into account terms up to and including $y^{4}$,eq.(22) reduces to:
\begin{equation}
	x_{1,2}=n_{1}+A[1\pm(1+\frac{y}{2}-\frac{y^{2}}{8}+\frac{y^{3}}{16}-\frac{5 y^{4}}{128})]
\end{equation}
This approximation  shows that the difference in number densities of two phases in equilibrium in a $LJ$ fluid depends on the parameters of both phases and (through the virial coefficients) on the parameters of the $LJ$ potential in the two phases.
The relative error $\delta$ of this expression compared to the exact result given by eq.(13) is $\delta\leq10^{-3}$ for $\left|y\right|\leq0.5$.

\subsection{The final result} 

The chemical potential of a $LJ$ fluid can also be expresed as [8]: 
\begin{equation}
	\mu=\mu_{ID}+2p_{ID}a_{2}
\end{equation} 
where $\mu_{ID}$ and $p_{ID}$ are the chemical potential and pressure of the ideal gas,and $a_{2}$ is the second virial coefficient. Imposing the condition of the equality of chemical potentials as a prerequisite for the phase equlibrium of two phases, it follows that
\begin{equation}
	\mu_{ID1}-\mu_{ID2}+ 2(p_{ID1}a_{21}-p_{ID2}a_{22})=0 
\end{equation}
Assume that the ideal gas behaves according to Maxwell-Boltzmann (MB) 
statistics. In that case, the pressure and chemical potential of the ideal gas are given by $\mu_{ID}=k_{B}T\ln(n\lambda^{3})$ and $p_{ID}=n k_{B}T$ [7]. Inserting these two expressions into eq.(25), and introducing $n_{1}-n_{2}=x$ finally leads to
\begin{equation}
\ln[(1+\frac{x}{n_{2}})(\frac{\lambda_{1}}{\lambda_{2}})^{3}]+2[(n_{2}+x)a_{21}-n_{2}a_{22}]=0
\end{equation}

Inserting the definition of $\lambda$ from eq.(4),one gets:
\begin{equation}
	\ln[(1+\frac{x}{n_{2}})\times (\frac{m_{2}}{m_{1}})^{3/2}]+2 [n_{2}(a_{21}-a_{22})+xa_{21}]=0
\end{equation}
which is obviously valid only for $n_{2} > 0$. 

This expression represents a theoretical form of the link between various parameters of two phases in equilibrium in a $LJ$ fluid. It contains $n$,$m$,$\lambda$,$\sigma$ and $\epsilon$ for each phase. Three of these parameters ($m$,$\epsilon$ and $\sigma$ ) are material dependent,while $\lambda$ contains a material parameter (the particle mass) and the temperature. In astronomical applications the number density depends on the nature of the object being studied.  To render it applicable, various terms in this equation have to be expressed in a more "practical form", in the sense of replacing the implicit dependence on various material (or object) parameters in this expression by an explicit formulation. 


\subsection{Possible cases} 

The behaviour of eq.(27) depends on the functions $A$,$y$ and $a_{2}$ which have been previously defined. 

In the case $y= - 1$, using eqs.(16) and (17),eq.(27) takes the following form:
\begin{equation}
\ln[1-2 a_{22} \lambda_{2}^{3}x]+ \frac{3}{2}\ln[\frac{m_{2}}{m_{1}}]+2(n_{1}a_{21}+A a_{22})=0
\end{equation}
under the condition $2 a_{22} \lambda_{2}^{3} x <1$. 

If $y=0$, there are two possibilities for $x$, given by eq.(18): $x=n_{1}+2A $ and $x=n_{1}$. In the first case, it can be shown that 
\begin{eqnarray}
	\ln[1+\frac{1}{n_{2}}\times(n_{1}+\frac{1}{a_{22}\lambda_{2}^{3}})]+\nonumber\\\frac{3}{2}\ln[\frac{m_{2}}{m_{1}}]+
2\times[(n_{1}+\frac{1}{a_{22}\lambda_{2}^{3}})a_{21}+\nonumber\\n_{2}\times(a_{21}-a_{22})]=0	
\end{eqnarray}
In the case $x=n_{1}$ one gets the following form of eq.(27): 
\begin{equation}
\ln[1+\frac{n_{1}}{n_{2}}]+2\times[n_{2}(a_{21}-a_{22})+n_{1} a_{21}]+\nonumber\\ 
\frac{3}{2}\ln[\frac{m_{2}}{m_{1}}]=0
\end{equation}
The difference $a_{21}-a_{22}$ is material dependent, through the values of $\epsilon$ and $\sigma$. 

An interesting limiting case corresponds to  $y$$\rightarrow$$+$ $\infty$. If $A$ and $n_{1}$ are finite, this case corresponds to $x_{1,2}=n_{1}-n_{2}\rightarrow -\infty$. Physically speaking,this means that 
the density of one of the phases is arbitrarily high and tends to infinity, while the other phase has a 
finite value of the density, which is remindful of the situation encountered in neutron stars.  
 

\section{THE PRACTICAL FORM}

The term containing the logarithm can be developed into series to give:
\begin{equation}
	\ln[(1+\frac{x}{n_{2}})]= \sum_{l=0}^{\infty}\frac{(-1)^{l}}{l+1}(\frac{x}{n_{2}})^{l+1}
\end{equation}
which is convergent for $\left|(x/n_{2})\right|<1$ (that is,$|n_{1}-n_{2}|<n_{2}$)and the explicit form of the function $x$ is given by eq.(12).Introducing the definition of $x$, this crietrion of convergence amounts to $n_{1}/n_{2}< 2$. The relative error of eq.(31) pushed to 3 terms compared to the exact result for the logarithm is $\delta\leq10^{-2}$ for $\left|(x/n_{2})\right|<0.5$. Using eqs.(5)-(7) for the second virial coefficient, it can be shown that
\begin{eqnarray}
	a_{21}-a_{22}= \frac{2}{3}\pi N_{A}\sigma_{1}^{3}\sum_{j=0}^{\infty}\frac{-2^{j+1/2}}{4 j!}\Gamma[\frac{2j-1}{4}]\nonumber\\
\left[(\beta\epsilon_{1})^{\frac{2j+1}{4}}-(\frac{\sigma_{2}}{\sigma_{1}})^{3}(\beta\epsilon_{2})^{\frac{2j+1}{4}}\right]
\end{eqnarray}

where $\beta = 1/k_{B}T$ and the second virial coefficient is approximately given by eq.(8)
Limiting the development in eq.(32) to terms up to and including $j=2$, it follows that
\begin{eqnarray}
a_{21}-a_{22}\cong3.62959N_{A}\sigma_{1}^{3}(\beta\epsilon_{1})^{1/4}\times [1-\nonumber\\
(\frac{\sigma_{2}}{\sigma_{1}})^{3}(\frac{\epsilon_{2}}{\epsilon_{1}})^{1/4}]-\nonumber\\
5.36938 N_{A}\sigma_{1}^{3}(\beta\epsilon_{1})^{3/4}[1-(\frac{\sigma_{2}}{\sigma_{1}})^{3}(\frac{\epsilon_{2}}{\epsilon_{1}})^{3/4}]-\nonumber\\	
1.81488 N_{A}\sigma_{1}^{3} (\beta\epsilon_{1})^{5/4} [1-(\frac{\sigma_{2}}{\sigma_{1}})^{3}(\frac{\epsilon_{2}}{\epsilon_{1}})^{5/4}]-\ldots
\end{eqnarray}
A practical problem related to the applicability of eq.(32) is the estimate of the relative error of this series,taking into account that there is no solution in closed analytical form for $a_{21}-a_{22}$. It can be estimated by choosing the parameters in eq.(32),and then determining the relative difference of the developement with $j$ and $j+1$ terms. Taking arbitrarily that $j=3$, $\sigma_{2}=2 \sigma_{1}$,$\beta\epsilon_{2}=0.05$,and that $\beta\epsilon_{1}\in(0.001,0.9)$, it follows that the relative error is $\delta\leq0.025$. 

Inserting eqs.(5)-(7),(13),(31) and (32) into eq.(27), it follows that
\begin{eqnarray}
\sum_{l=0}^{\infty}\frac{(-1)^l}{l+1}[\frac{n_{1}+A[1\pm(1+y)^{1/2}]}{n_{2}}]^{l+1}+ \nonumber\\
(3/2)\ln(m_{2}/m_{1})+2\times n_{2}\times\frac{2}{3}\pi N_{A}\sigma_{1}^{3}\nonumber\\
\times \sum_{j=0}^{\infty}\frac{-2^{j+1/2}}{4 j!}\Gamma[\frac{2j-1}{4}]\nonumber\\
\times \left[(\beta\epsilon_{1})^{\frac{2j+1}{4}}-(\frac{\sigma_{2}}{\sigma_{1}})^{3}(\beta\epsilon_{2})^{\frac{2j+1}{4}}\right]\nonumber\\
+2\times(n_{1}+ A[1\pm(1+y)^{1/2}])\times\frac{2\pi N_{A}\sigma_{1}^{3}}{3}\nonumber\\
\sum_{j=0}^{\infty}\frac{-2^{j+(1/2)}}{4 j!}\nonumber\\
\times \Gamma[(2j-1)/4](\beta\epsilon_{1})^{(2j+1)/4}\nonumber\\=0
\end{eqnarray}

This equation links various parameters of two phases in equilibrium of a $LJ$ fluid,and it can be explicitely written with an arbitrary number of terms. Taking only the terms with $j\in(0,3)$ and $l\in(0,3)$ in various sums in eq.(34),expanding out all products and integer powers leads to the following result
\begin{eqnarray}
(3/2)\ln[m_{2}/m_{1}]+\frac{x}{n_{2}}\times[1-(1/2)\frac{x}{n_{2}}+\nonumber\\
(1/3)(\frac{x}{n_{2}})^{2}-(1/4)(\frac{x}{n_{2}})^{3}]+N_{A}n_{2}\sigma_{1}^{3}(1+\frac{x}{n_{2}})\nonumber\\
\times[7.25918(\beta\epsilon_{1})^{1/4}-3.62959(\beta\epsilon_{1})^{5/4}-\nonumber\\
10.7388 (\beta\epsilon_{1})^{3/4}-1.7898(\beta\epsilon_{1})^{7/4}]\nonumber\\
+N_{A}n_{2}\sigma_{2}^{3}(\beta\epsilon_{2})^{1/4}[10.7388(\beta\epsilon_{2})^{1/2}-\nonumber\\
7.25918(1-(1/2)\beta\epsilon_{2})]=0
\end{eqnarray}
where $x$ has been defined in eq.(13).  

A simplifying case of physical interest is $\beta\rightarrow 0$,which corresponds to high temperature,when several terms in eq.(35) tend to zero.Another interesting simplification corresponds to $x=0$. 
The last two expressions can be further simplified by assuming that $m_{2}=m_{1}$,which implies that also $\lambda_{2}=\lambda_{1}$

\section{POSSIBLE APPLICATIONS IN ASTROPHYSICS} 

Equation $(34)$ links two kinds of parameters,and it can be applied in two ways.It contains the number densities and the thermal wavelengths of the two phases, but it also implicitely contains the second virial coefficients in the two phases. This expression can be applied to several kinds of systems: cold diffuse clouds in our galaxy,the interiors and the atmospheres of the giant planets or icy satellites, and possibly also to neutron stars. 

A cold cloud sufficiently far from any star or other source of radiation certainly contains neutral atoms and/or molecules,which are known to interact via the $LJ$ potential. Immagine that such a cloud contains two phases in equilibrium,which means that their pressure,temperature and chemical potential are equal. Take,for example,that phase $1$ is the gas at the outer edge of such a cloud,and that phase $2$ is the gas some distance below the top. If the particle number density in one of the phases,$n_{1}$, can be measured, eq.(34) provides a theoretical answer to the following question: What is the value of $n_{2}$ so that the cloud as a whole remains in phase equilibrium? A possible consequence of the impairment of phase equilibrium can be the onset of star formation.
  
Globally speaking, giant planets and the icy satellites consist of fluids and deep in their interiors they may contain a dense central core.
The fluid envelopes are made up mostly of hydrogen and helium, with (possible) small additions of heavier chemical elements. The cores are presumed to consist of combinations of refractory elements. The exact combination of elements making up a planetary core can not be determined directly. It can only be checked indirectly, by constructing a model of the interior structure assuming some composition of the core,and then comparing the observable consequences of a model with observed data. 

In the outer layers of the atmospheres of the giant planets matter is neutral,and can accordingly be described by a $LJ$ potential, while in the deeper regions ionization, strong ion coupling and electron degeneracy become important [11]. Modelling the interior structure of such a system is a complex task in statistical physics; it demands the knowledge of the chemical composition and the equation of state. The chemical composition can be determined from observation only for the uppermost layer of a giant planet or a satellite. Regardless of the exact form of the equation of state, pressure and density of matter increase with depth. Matter making up a planet or satellite consists of atoms and molecules, which under  increasing pressure become excited and finally ionised.

Taking these considerations into account, how can eq.$(34)$ be applied? Material parameters of the highest levels of the atmosphere can be determined from observation.In order to apply the calculations discussed in this paper,it has to be assumed that that the $LJ$ potential is applicable at and/or near the upper layers of the atmosphere of the object,and that the object is isothermal,at least until a certain limiting depth. The next step is to determine,within any particular theoretical framework, the number of phases which exist within the isothermal region of the object being considered,and assume that these phases are in equilibrium. When this number is known,eq.$(34)$ gives the possibility to calculate the values of the second virial coefficient $a_{22}$ and of the thermal wavelength $\lambda_{2}$ in the phase immediately below the surface. It becomes possible by repeating this procedure to "follow" changes of the second virial coefficient and the thermal wavelength with depth below the surface.

Interesting conclusions can be drawn concerning the time evolution of the equilibrium in a $LJ$ fluid. Immagine that in a given moment of time one of the parameters in eq.$(34)$ changes. A change easiest to envisage is the change of the particle mass in one of the phases. One would expect that in such a situation the equilibrium would be disturbed and that the system (or the part of it which was considered) would start evolving in time. There is however another possibility: that some other parameter (or parameters) changes in such a way that the equilibrium is maintained. Details will be discussed separately. 

\section{CONCLUSIONS}

In this paper we have analyzed the phase equilibrium in a Lennard Jones fluid. The starting points were the basic conditions for phase equilibrium, well known in statistical physics,and the recent expression for the chemical potential of a Lennard Jones  fluid. The result of the calculation is an expression for the condition for phase equilibrium, given by eq.(27) in its most general form,or by eq.(34) for the case of a Lennard Jones fluid. In deriving eq.(34) it was assumed that the ideal gas contribution to eq.(25) is governed by Maxwell- Boltzmann statistics. It would be interesting to repeat the same calculation, but with another form of statistics. Several possible applications in astrophysics were indicated, including one unexpected - the neutron stars.  
    

\section{Acknowledgement}
This paper was prepared within the research project $174031$ financed by the Ministry of Education and Science of Serbia.The author is grateful to the referee for useful comments about this manuscript.



{

{\ }








{}













{




\end{document}